\documentclass[prb,twocolumn,superscriptaddress,groupedaddress,amsmath,amssymb,showpacs]{revtex4}
\usepackage{graphicx}
\usepackage{dcolumn}
\usepackage{bm}
\usepackage{color}

\begin{document}
\preprint{APS/123-QED}

\title{Time domain optical manipulation of exciton and nuclear spin in a single self-assembled quantum dot}
\author{H. Sasakura}
\affiliation{CREST, Japan Science and Technology Agency, Kawaguchi 332-0012, Japan}

\author {S. Adachi}
\affiliation{CREST, Japan Science and Technology Agency, Kawaguchi 332-0012, Japan}
\affiliation{Department of Applied Physics, Hokkaido University, N13, W8, Kitaku, Sapporo 
060-8628, Japan}

\author {S. Muto}
\affiliation{CREST, Japan Science and Technology Agency, Kawaguchi 332-0012, Japan}
\affiliation{Department of Applied Physics, Hokkaido University, N13, W8, Kitaku, Sapporo 
060-8628, Japan}

\author {S. Hirose}
\affiliation {Fujitsu Labs Ltd., 10-1 Morinosato-Wakamiya, Atsugi 243-0197, Japan}

\author {H. Z. Song}
\affiliation {Fujitsu Limited, 10-1 Morinosato-Wakamiya, Atsugi 243-0197, Japan}
\author {M. Takatsu}
\affiliation {Fujitsu Limited, 10-1 Morinosato-Wakamiya, Atsugi 243-0197, Japan}

\date{\today}

\begin{abstract}
We have demonstrated experimentally the manipulation of exciton and nuclear spins in a single self-assembled In$_{0.75}$Al$_{0.25}$As/Al$_{0.3}$Ga$_{0.7}$As quantum dot. The oscillation of exciton and nuclear spin polarizations were clearly observed. The switching of the emissions in Zeeman split pair indicates that the exciton pair with opposite spins was created coherently via the continuum states and that we can control the electron and nuclear spin polarizations only by changing the delay time of the cross-linearly-polarized pulses. These suggest the high potentiality of electron and nuclear spin manipulation in a single QD via the continuum state.

\end{abstract}

\maketitle


Recently much attention has been paid to electron spins in semiconductor quantum dots (QDs). The electron spin in QD is proposed to be used for quantum information processing such as quantum computation \cite{Loss98, Sasakura03} and media conversion between photon qubit and electron spin qubit \cite {Muto05}. One of the advantages for electron spins in QDs is the controlability by using the optical pulse with the band gap energy. Electron spin manipulation by the $\pi$-pulse technique were already shown  in a single QD \cite{Duncan03,Villas05,Zrenner02,Stievater01,Kamada01}. It requires the rigorous handling to tune the pump pulse area and energy (which resonate with the energy difference between the target electron spin states for the spin-flip operation in a single QD and the spin transfer to another dot). Also, some proposals of the spin manipulation in a $\Lambda $-type transition scheme uses the multi-step processes via the discrete or virtual state in a QD \cite{Imamo99, Biolatti00,Fliss04}. However, among the enormous amount of QDs, it is really difficult to find a suitable single QD that is appropriate as the intermediate state. Also, the large excitation power was the hurdle for the virtual intermediate state. Stimulated Raman adiabatic passage (STIRAP) is a powerful tool for the complete population transfer between the target states by the coherent interaction via an intermediate state \cite {Vitanov01}. Recently, the experimental demonstration of coherent population transfer in helium atoms via continuum states using STIRAP has been shown \cite{Peters05}. Using the continuum state in QD as the intermediate state, the large experimental advantage is gained . The continuum state always has the connecting path between Zeeman-split exciton pair with opposite spins under the magnetic field. Although the continuum state has the faster relaxation process than the discrete state in QD \cite{Toda98}, it should not be an obstacle because STIRAP process does not create the  population in the continuum state. One important issue is that the continuum state in a QD has the practical oscillator strength for multi-pulse processes. Here, in a single InAlAs QD, we report the effect of the continuum state in the excitation process where 
two coherent visible light pulses are used. 
The coherent polarization oscillation of exciton and nuclear spins in single QD are clearly observed as a function of the temporal delay between two optical pulses.

\begin{figure}[t]
\includegraphics[width=60mm]{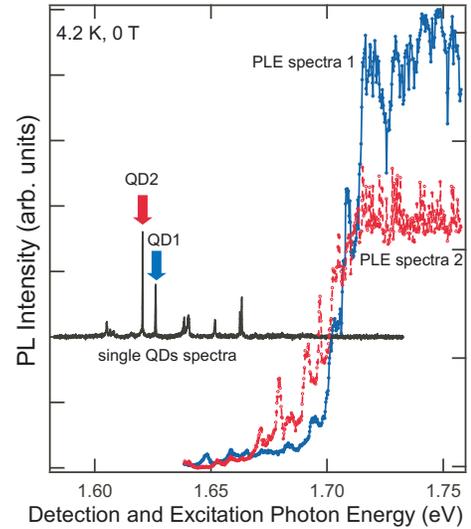}
\caption{(color) Single QD photoluminescence (black line) under He-Ne laser excitation (633 nm). Solid (blue) and dashed (red) lines indicate PLE spectra of QD1 and QD2 under cw-Ti:S laser excitation, respectively. Both spectra were taken under zero magnetic field at 4.2 K. Two peaks labeled as QD1 and QD2 are assigned to exciton emissions from the different QDs because of no correlation between PLE spectra 1 and 2.}
\label{Fig1}
\end{figure}
The InAlAs QD were grown on a AlGaAs (120 nm)/GaAs (300 nm) buffer layer on undoped (100) GaAs substrates by a molecular beam epitaxy by using a RIBER-MBE32. The growth temperature was 620 $^\circ$C for the buffer layer and 500 $^\circ$C for the InAlAs QDs. The substrate temperature was lowered gradually during the growth of the buffer layers. The InAlAs QDs were grown at a rather low growth rate of 1.5 $\times $ $10^{-3}$ ML/s. For stabilizing the InAlAs QDs, post-annealing of 90-sec. was performed. Then the substrate temperature was gradually increased to 590 $^\circ $C during AlGaAs capping layer (100 nm) was grown. Small mesa structures were fabricated by electron-beam lithography and wet chemical etching to isolate a single QD. 

Figure 1 shows conventional photoluminescence (PL) spectrum and photoluminescence excitation (PLE) spectra of single QD. He-Ne laser was an excitation source in the PL measurement and a continuous wave (CW) Ti:sapphire (TIS) laser were scanned 1.654 $\sim$1.758 eV in the PLE measurement. Two main emissions (QD1 and QD2 in Fig. 1) have the energies 1.6263 eV (FWHM 294 $\mu$eV) and 1.6206 eV (FWHM 400 $\mu$eV), respectively. Since there is no correlation between PLE spectra 1 and 2, those two PL peaks (QD1 and QD2) can be assigned to the emissions from different QDs. Both PL peaks are located on the low energy tails of the continuum states as shown in PLE spectra.
\begin{figure}[h]
\includegraphics[width=75mm]{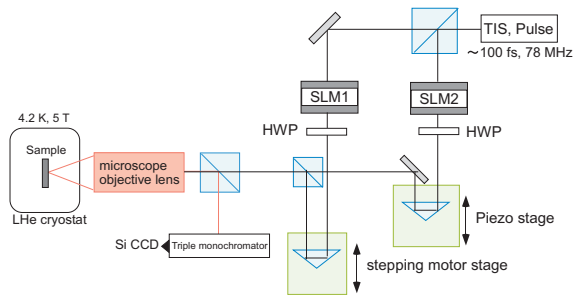}
\caption{Schematic experimental setup based on a Mach-Zehender interferometer includes two dispersion-free 4-f optical systems. The time delay and polarization between two pump pulses are controlled by the piezo stage and HWP angle.}
\label{Fig2} 
\end{figure}

The schematic experimental setup is shown in Fig. 2. 
A mode-locked TIS laser with a pulse reputation rate of 76 MHz was used as a light source. A sequence of the pump pulses were produced by passing through a Mach-Zehender interferometer. The 100-fs laser pulse was divided into two pump pulses with a controlled time delay using a piezo stage with $5$ nm resolution. The pump pulse pair was slightly shaped spectrally by two dispersion-free 4-f optical systems in conjunction with spatial light modulators (SLMs) which has $350$ $\mu$eV resolution for cutting off the laser spectra tail. Then two pump pulses were combined concentrically and were focused on the sample surface by an objective lens. The single QD emissions collected by the same objective lens were dispersed by a triple grating spectrometer ($f=0.64$ m) and were detected with a LN$_{2}$-cooled Si-charge-coupled device (CCD) camera which has 12 $\mu$eV resolution and the energies of the emission peaks was determined on the order of 3 $\mu$ eV by the spectral fitting. A mode-locked TIS laser was tuned to $\sim 1.7$ eV for the excitation of the bottom of the continuum states determined by the PLE measurement.
 The continuum states are "half-localized" in QDs, in the sense that it consists of electron and hole states, one of which is confined in the QD. Therefore the carriers are localized in the QD after the energy relaxation.
 The sample was held in a LHe cryostat and was kept at 4.2 K. A magnetic field of 5 T along the growth direction was applied. The pump pulse pair were either cross-linear or co-linear by adjusting the two half-wave plates (HWP) after SLMs.
\begin{figure}[h]
\includegraphics[width=85mm]{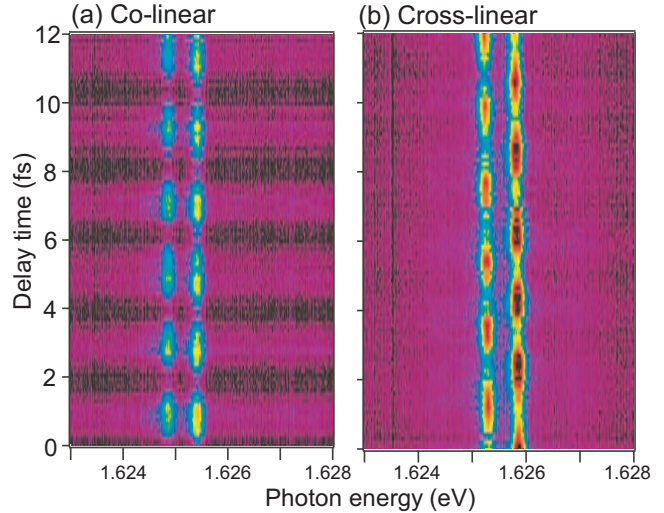}
\caption{(color) Contour plots (CCD image) of the Zeeman pair as a function of the pulse delay. Two pulse was introduced to sample with co-linear (a) or cross-linear (b) polarization by the changing HWP angle.}
\label{Fig3} 
\end{figure}
\begin{figure}[h]
\includegraphics[width=85mm]{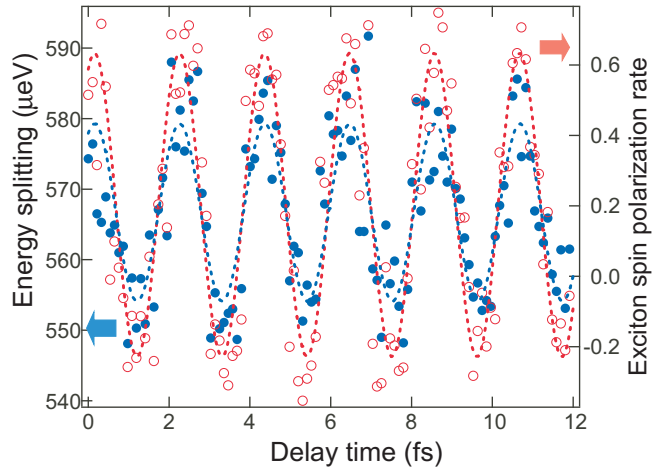}
\caption{(color) The oscillation of the energy splitting (solid circles) and the exciton polarization rate (open circles) as a function of the pulse delay in cross-linear polarization setup. Two dash lines (blue and red) were fitted with sine function.}
\label{Fig4} 
\end{figure}

In Fig.~\ref{Fig3}, we show the contour plots of the exciton Zeeman spectra of QD1 as a function of 
the pulse delay. With the pump pulses of co-linear polarization, the well-known coherent oscillation of exciton ground state in single QD was observed~\cite{Bonadeo98,Toda00,Besombes03}. Precisely, the time-integrated PL of the Zeeman-split exciton pair oscillates {\it in-phase} as a function of the pulse delay and the oscillation period corresponds to the energy of the excitation state (Fig.~\ref{Fig3}(a)). The oscillation amplitude of the exciton ground state emission is convergent within the overlapping time of the pump pulse pair, which indicates that the coherence of continuum state is limited roughly to the inverse of the energy bandwidth of the excitation state. 
In the case of cross-linear polarizations of the pump pair, the two Zeeman split emissions oscillate {\it anti-phase} with the same frequency as that in co-linear polarization. The phase difference between the Zeeman pair originates from the 
the interference between the continuum states generated by orthogonally-polarized optical pulses (Fig.~\ref{Fig3}(b)).
The modulated coherent exciton spin polarization indicates that the exciton pair with opposite spins is created through the continuum states to an extent enough for practical electron spin manipulation in single QD.

Figure~\ref{Fig4} shows the oscillations of the exciton spin polarization and the energy splitting between Zeeman pair as a function of delay time in the pump pair of cross-linear polarizations (Fig.~\ref{Fig3}(b)). We can estimate the exciton spin polarization by the degree of circular polarization $(I_{+}-I_{-})/(I_{+}+I_{-})$ from the observed spectra. The peak energies of the Zeeman pair are estimated by the spectra fitting. 
The splitting energy of Zeeman pair changes up to $\sim$45 $\mu$eV with varying the delay time. This energy change synchronizes the exciton spin polarization and is considered to be induced by "Overhauser field"~\cite{Overhauser53}. The Overhauser field originates from the nuclear spin polarization that can be created by the Fermi contact interaction via spin flip-flop process with the electron spin in an exciton\cite{OptOrientation}. The direction of Overhauser field is decided by the electron spin direction and the sign of electron g-factor, and therefore can be controlled by the polarization of the excitation pulse~\cite{Yokoi05}. In this experiment, the polarization change of the electron spin is created by the delay time between two cross-linearly polarized pump pair and the direction of Overhauser field changes following the exciton spin polarization. Therefore, the splitting energy of Zeeman pair changes with varying the delay time. 
These data shows that we can control the electron and nuclear spin polarization only by changing the time delay of two cross-linearly polarized pulses.

In summary, we have experimentally demonstrated the temporal domain optical manipulation of the electron and nuclear spins in a single QD through the excitation of the continuum states originating from QD.
These results show the possibility to use the QD continuum state as the intermediate state in multi-pulse process of the control operation. We have not yet done the STIRAP for the electron spin manipulation. However, the results show that the continuum state half-localized in QD has enough oscillator strength to be used as intermediate state for the $\Lambda$ transition.

\end{document}